# INVESTIGATING THE NATURE OF INTANGIBLE BRAIN-MACHINE INTERACTION

FOTINI PALLIKARI

*Solid State Physics Depatment, Faculty of Physics, National and Kapodistrian University of Athens, Greece*

**Abstract.** The hypothesis that direct human intention can modulate the concurrent outcomes of a stochastic process has been under test for over 30 years, surrounded by inconclusive evidence and a great amount of ambiguity. An increased interest has recently and prematurely risen concerning possible scientific applications of such effects. The amassed information around the purported phenomena, as presented in a meta-analysis [1], is thoroughly re-examined here through the application of several recognised mathematical tools. It is shown that there exists no validation of the hypothesis of brain-machine interaction precisely in absence of interface devices. It is concluded that any evidence in favour of the alleged intangible brain-machine effects, must have resulted from unintended errors during data collection and treatment, known as "the experimenter expectancy effect".

Keywords: Random number generators, mental process, Markov process, rescaled range analysis, experimenter expectancy effect.

## INTRODUCTION

A large-scale experiment aiming to replicate claims of ostensive interaction between direct human intention and the statistical outcomes of random processes resulted in the refutation of the alleged hypothesis [2]. Prompted by the replication failure, a meta-analysis was carried out [1] involving all studies that had been performed over the last 35 years to examine the effects of direct human intention on the concurrent output of a true random number generator (RNG). The meta-analysis included all those of the reported studies whose outcome could be converted into binary form. The result of this meta-analysis was ambiguous; an overall weak effect was observed, that was nevertheless attributed to flaws in the associated database introduced by publication bias.

The scientific importance of such investigation is obvious, considering the potential hazardous interference human operators could perpetrate on the operation of sensitive equipment. The tremendous potential in science and technology of brain-machine interface devices has already been realised by several applications [3, 4]. The benefits of possible control of a physical process through our thoughts directly, i.e. without the intervention of an interface device, would obviously be far greater.

Scientists have already attempted to explore the direct influence of human thoughts on a physical system, the quantum state of photons [5], reporting positive results. Concerns have been raised, however, about the treatment of data and the appropriateness of the applied experimental arrangement to demonstrate the claimed effects [6], which makes similar endeavours [7] equally unconvincing. Surrounded by such ambiguities the question of whether direct human intention can modulate a stochastic process remains critically open.

The current work aims to remove these ambiguities by carefully evaluating the available evidence in order to conclusively identify the origin of the mechanism underlying the elusive brain-machine interaction. For brevity purposes this type of interaction will be dubbed here as the Intangible Mind-Machine Interaction (IMMI).

To address this matter, the reported meta-analysis data [1] will be treated by several mathematical tools, as follows. Data will be represented by funnel plots to evaluate the aptness of the IMMI database to produce reliable statistical information. Furthermore, the available experimental records will be treated as the outcomes of a Markovian IMMI source operating at fixed transition probabilities to assess the likelihood of existence of such mechanism. The nature of persistent correlations revealed by the application of the above mathematical treatments will be further explored by applying the rescaled range analysis (R/S) on time series generated by the records of these true RNG processes.

## METHODS

### *Funnel plots: The IMMI RNG database*

The true RNG processes in the meta-analysis studies [1] generate the equivalent of binary sequences; sequences of 1's and 0's, or hits and misses. The final outcome of an IMMI study is expressed as effect size, pi, in terms of $P_{obs}$, the raw proportion of hits in the study as

$$pi = \frac{P_{obs}(\kappa-1)}{1+P_{obs}(\kappa-2)} \overset{\kappa=2}{\equiv} P_{obs}, \quad 0 \le pi \le 1 \qquad (1)$$

The parameter κ in equation (1) represents the number of alternative choices available in the RNG IMMI test. For instance, in the case of electronic RNG, κ = 2, so that the effect size pi becomes equal to $P_{obs}$. The result of the study is also expressed as z-score and estimated via the parameter pi through

$$z = \frac{pi - 0,5}{se_{pi}} \qquad (2)$$

| RNG Database | N | $\wp$ | $(\overline{pi})$ | $\overline{pi}_{overall}$ |
|---|---|---|---|---|
| IMMI | 380 | 0,500 | 0,510 (0,002) | 0,503 (0,002) |
| Control | 137 | 0,500 | 0,495 (0,001) | |

Table 1. Characteristic parameters associated with funnel plots. N: The size of study; $\overline{pi}$: Statistical average of all effect sizes included in the database; $\wp$: The graphically estimated most representative effect size, pi, to which the pi values converge at very large study sizes, N; $\overline{pi}_{overall} = (\overline{pi}_{IMMI} + \overline{pi}_{control})/2$, the combined average of IMMI and control $\overline{pi}$ values. Standard errors of estimates are shown in brackets. The standard error of $\overline{pi}_{overall}$ was estimated assuming that both databases originate from the same population of true RNG data.

The standard error of pi in equation (2), $se_{(pi)}$, in a study of N collected bits, is estimated through

$$se_{pi} = \frac{pi(1-pi)}{\sqrt{NP_{obs}(1-P_{obs})}} \quad (3)$$

The theoretically expected proportion of hits in a true RNG IMMI study is $P_{th}=1/\kappa = 0{,}5$. If direct human intention could influence the statistics of the random process, according to the IMMI hypothesis, the proportion of hits in a test should appear significantly shifted above the theoretically expected value $P_{th}$ beyond statistical error. Validating the IMMI hypothesis, however, requires the collective evidence from all similar scores generated by true RNG processes, after correcting possible flaws such as the publication bias, which results from the reluctance of experimenters to report their experiments (often when results contradict the investigated hypothesis).

The presence of biases in a database can be visually observed by constructing the so-called funnel plot [8]. In one of the possible forms of this graphical representation of a database, the size of study N is plotted against the associated effect size pi. Provided the database is adequately large and free of biases, the plot is expected to be symmetric and regressing to the most representative value of effect size in the database $\wp$, so that its shape will resemble the shape of an inverted funnel.

The funnel plot of true IMMI RNG data, figure 1, is a rich source of information about the associated database. For instance, the most representative IMMI effect size is $\wp_{IMMI} = 0{,}5$. This is the effect size to which the IMMI funnel plot converges at very large size studies N ($> 10^6$ bits) and whose location is shown in figure 1 by the position of the dashed vertical line. This result is exactly what would be expected in a process where the IMMI hypothesis is not manifest.

Yet, the IMMI database of the 380 studies included in the meta-analysis [1] is clearly marked by publication bias as seen in figure 1. The funnel plot is not adequately symmetric as it should, being densely populated in the region $0{,}50 < pi < 0{,}65$ and N< $10^5$ bits, whereas it is void of data at effect sizes $0{,}35 < pi < 0{,}50$. Due to this asymmetry in the way the IMMI data scatter about $\wp_{IMMI} = 0{,}5$, more prominently around the base of the funnel plot, the statistical average of the 380 IMMI effect sizes is $(\overline{pi})_{IMMI} = 0{,}510$ (standard error 0,002), indicated in figure 1 by the vertical dash-dotted line; The IMMI statistical average $(\overline{pi})_{IMMI}$ does not coincide, as it should, with the most representative effect size $\wp_{IMMI} = 0{,}5$ and cannot provide a reliable statistical estimate of the investigated effect.

Even more pertinent is an additional type of bias present in the IMMI database; one which challenges the presumed random variability of records in the reported IMMI database. The scatter of IMMI data about $\wp_{IMMI} = 0{,}5$ is also wider than would be expected in a database of independent experimental records. To appreciate this defect, the scatter of random data about their statistical average is plotted being confined within the confidence intervals $pi = f(N, z_0)$, equation (4), [9]. The parameter $z_0$ in equation (4) determines the level of statistical significance adopted before a tested hypothesis is assessed to be true or false on the basis of existing experimental data.

$$pi = 0{,}5 \pm z_0 \cdot \frac{0{,}5}{\sqrt{N}} \overset{z_0=1{,}96}{=} 0{,}5 \pm \frac{0{,}98}{\sqrt{N}} \quad (4)$$

$$N = \frac{0{,}96}{(pi - 0{,}5)^2} \quad (5)$$

For instance, for $z_0$ = 1,96 the two confidence interval curves drawn on the funnel plot according to equation (5) constitute the range of 95% statistical confidence of true random data. It is expected that the 95% of plotted random data in a large enough database are enveloped by the two blue dashed curves in figure 1. Assuming random variability of the 380 of IMMI data, we would expect to observe an average of 5% of them located on and outside the two 95% confidence interval curves. Yet, a much larger number of data points are plotted in that region, indicating a broadening of their scatter as compared to the scatter of random data.

To better represent the confidence interval curves that fit the IMMI data, which do not exhibit random variability, a more general form of equation (4) can be invoked, equation (6), on the basis of the Markovian model that will be discussed in the following section. The parameter $z_0$ is now multiplied by the variance factor V=2,21 as if the z-value has increased from $z_0$ = 1,96 to 4,33, so that the 95% of the present IMMI data points are adequately enveloped by the confidence interval, table 2.

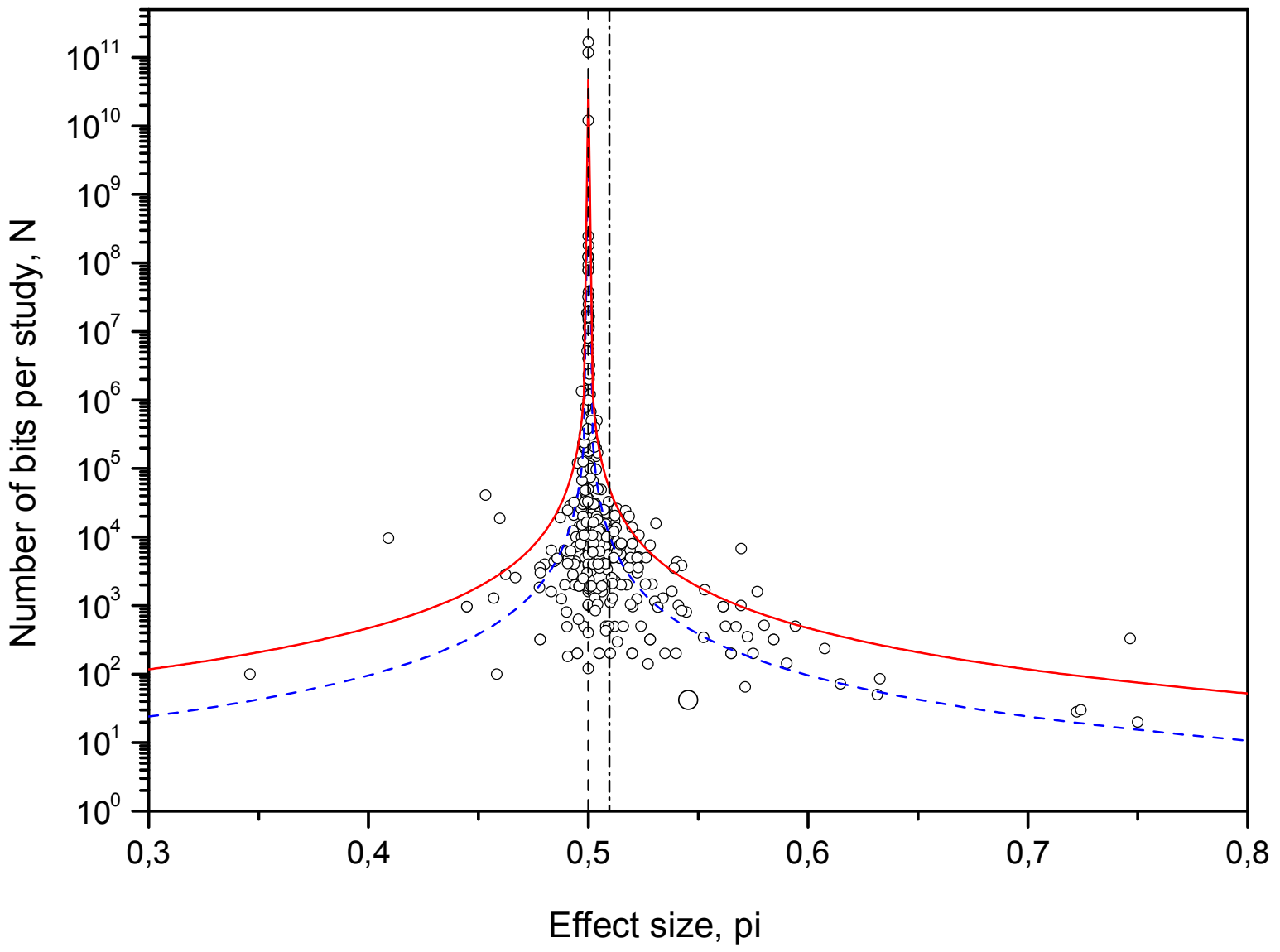

Fig. 1. Funnel plot of scores, pi, in IMMI studies reported in [1]. *Blue dashed curves*: confidence interval enveloping the 95% of randomly distributed data ($z_0$ =1,96; equations 4 and 5). *Red solid curves*: broadened confidence interval to envelope the 95% of the meta-analysis data. *Dashed vertical line*: positioned at the value of pi to which the funnel plot converges ( $\wp = 50\%$ ). *Dash-dotted vertical line*: positioned at the simple statistical average of all plotted data ( $\overline{pi} = 0{,}510$ ; standard error=0,002). Standard errors of pi values are not displayed on the graph for clarity.

Such graphical estimation of the broadening of variance although approximate, it is nonetheless quite telling of the presence of bias, a persistent type of data correlation in the database. In other words, the reported effect effect sizes of IMMI studies, pi, are correlated and not randomly distributed records of independent experiments. Exploring the nature of these correlations at the level of bit generation as well as at the level of published experimental scores, will be the subject of the next sections. The analysis will invoke the Markovian approach to represent the IMMI mechanism, as being a universal Markovian process operating at fixed transition probabilities.

| **Markovian process** | **V** | **p** | **$C_1$** |
|---|---|---|---|
| IMMI | 2,21 | 0,83 | 0,66 |
| Memoryless | 1,00 | 0,50 | 0 |

Table 2. Graphically estimated parasmeters determining the biased Markovian IMMI RNG process, for which the self-transition probabilities are $p_{11} = p_{00} = p$, as compared to those of a true memoryless process. V: variance factor, equation (8c); $C_1$: Markovian first neighbour correlation coefficient, equation (9).

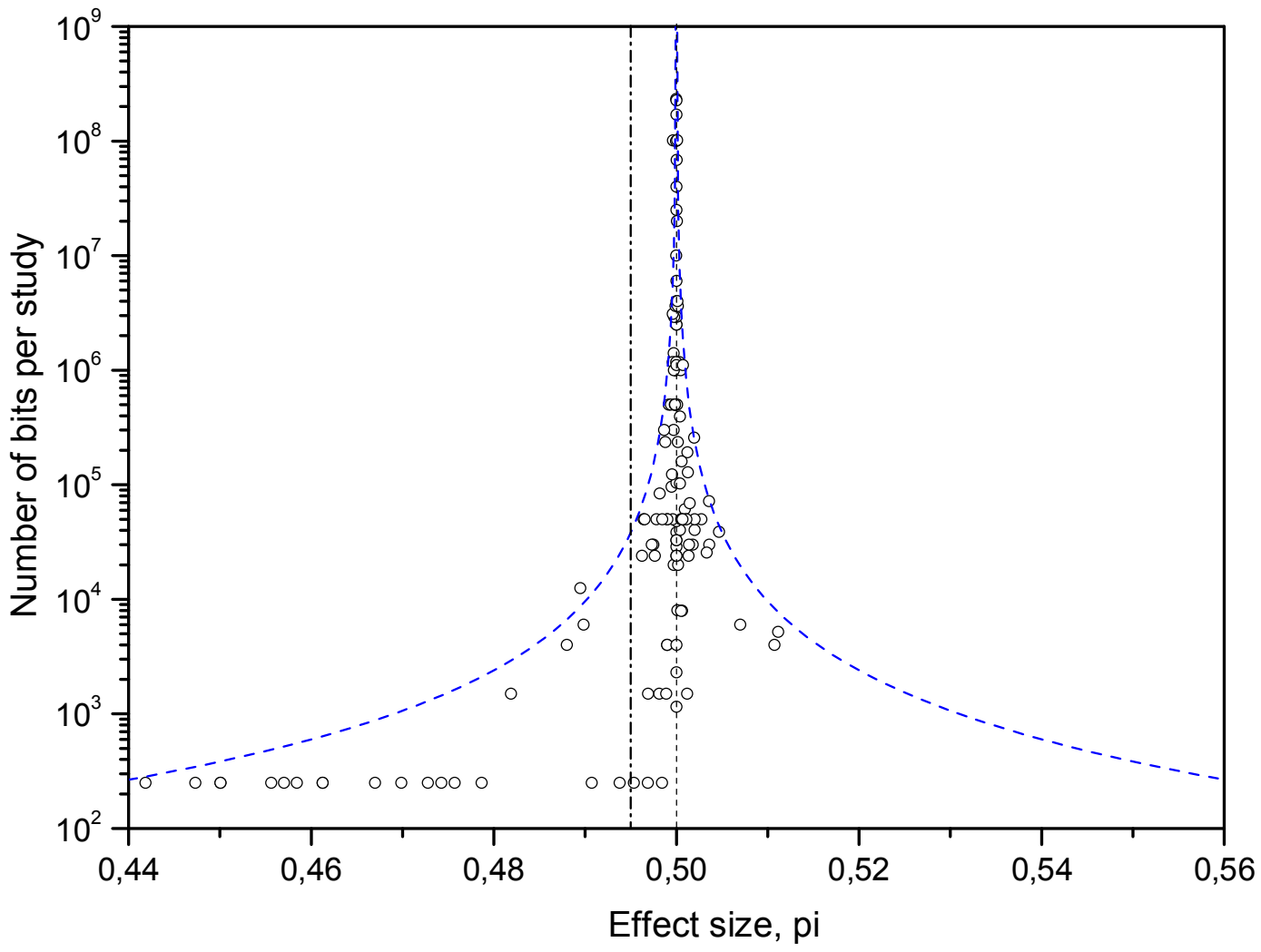

Fig. 2. Funnel plot of effect sizes obtained in control tests reported in the meta-analysis of [1]. *Blue dashed curves*: 95% confidence interval. *Dashed vertical line*: positioned at the most representative effect size, to which the funnel plot converges and which coincides with the theoretically expected statistical average of a random binary process $\wp_{\text{control}}$= 0,50. *Dash-dotted vertical line*: positioned at the simple statistical average of control effect sizes $(\overline{pi})_{\text{control}} = 0{,}495$; standard error=0,001. The standard errors of pi values are not displayed on the graph for clarity.

*Funnel plots: The control RNG database*

The meta-analysis [1] has also provided records of control data generated by the same RNG processes as randomness checks to ensure that the devices were functioning properly. Before introducing the Markovian model, it is worth evaluating the information provided by the funnel plot of control data.

The funnel plot of only 137 such control scores reported in [1] is shown in figure 2. The plot converges to 50%, same as the funnel plot of the IMMI data does. The 95% confidence interval curves, the blue dashed lines in figure 2, adequately envelope the 95% of data points. The 95% confidence interval in this case is considered “adequate” in view of the distorting presence of strong publication bias and the relatively smaller size of the database, in spite of an apparent weak narrowing of the data spread,.

Therefore, the random variability of the reported control data can be adequately confirmed as far as the spread of data, regardless of the clear presence of publication bias in the database. The presence of publication bias on the funnel plot is now visible at the lower right hand side ($N < 10^4$), largely at effect sizes pi $> 50\%$. Presumably, these voids are due to the reluctance of experimenters to report effect sizes that occur within a range falsely considered as being unsuitable for control RNG experimental records. Still, control data can naturally display such effect sizes without challenging the efficiency of the true RNG

| | **RNG-IMMI** | **RNG-Control** | **Random Sequences** | | **RNG-Randomized IMMI** | **RNG-Randomized Control** |
|---|---|---|---|---|---|---|
| **N** | 380 | 137 | 380 | 137 | 380 | 137 |
| **H** | 0,70 (0,05) | 0,68 (0,03) | 0,56 (0,03) | 0,64 (0,07) | 0,55 (0,02) | 0,57 (0,05) |
| $\mathbf{C_H}$ | 0,32 (0,09) | 0,28 (0,05) | 0,09 (0,05) | 0,21 (0,02) | 0,07 (0,03) | 0,10 (0,08) |

Table 3. Characteristic parameters in rescaled range analysis, R/S, performed on time series consisted of IMMI and control data generated by true RNG processes, as well as on random sequences generated online [22] for reference. N: Number of records in the time series; H: Hurst exponents estimated by linear regression through equation (13). $C_H$: correlation coefficient, equation (14). RNG-Randomized: The average parameters estimated from ten randomized RNG time series. The accuracy of estimates in brackets denotes standard error. Standard error is obtained through linear regression in the case of the true RNG sequences and by averaging over ten sequences for the random and RNG-randomised sequences.

process.

Yet again, the problem of publication bias is stronger at the lower part of the funnel plot of control data, i.e. the region of the small size studies. The publication bias present in the control data brings the overall statistical average at $(\overline{\text{pi}})_{\text{control}} = 0{,}495$ (standard error = 0,001), i.e. below the value of the most representative effect size on the plot: $\wp_{\text{control}} = 50\%$, whose location is shown in figure 2 by the vertical dash dotted line and also presented in table 1. Past attempts to correct statistical distortions introduced by publication bias in a database have involved the use of various corrective statistical tools, which are not generally acceptable [10].

The net result of the publication bias observed in either control or the IMMI databases is to shift their statistical average away from the most representative effect size, $\wp = 50\%$ and chance expectation in opposite directions, table 1. If the two databases are merged, which is reasonable doing since both IMMI and control databases have been generated by the same RNG processes, then the contrasting consequence of publication bias brings the overall statistical average to $\overline{\text{pi}}_{\text{overall}} = 0{,}503$ (standard error = 0,002), table 1. This result falls within the statistically expected in a true random RNG process. The merging of data has on the one hand increased the size of database to N=517, improving the statistical power of the estimate and on the other hand has statistically restored the average effect size to the value expected in a true RNG process, $\wp_{IMMI} = \wp_{control} \equiv 50\%$.

***The Markovian IMMI RNG process***

The bits collected across experiments in the true RNG IMMI meta-analysis [1], will be treated as the outcomes of one and the same hypothetical universal IMMI process. Each IMMI study has generated a separate number of bits from which the associated experimental effect size was estimated. In that sense, it is supposed that the same IMMI mechanism is in operation every time such bits are generated. In the current treatment, this imaginary process will obey the same Markovian rules through which all the data in the meta-analysis is supposed that have been generated.

The Markovian source will operate as follows. With every bit generated the true RNG is brought to a binary state, either 1 or 0. The state 1 is the code for RNG outcomes in the prestated intented direction and it is 0 otherwise. Being a Markovian source, each new state of the true RNG IMMI process depends on its previous state. There will be, therefore, four state transition probabilities, $p_{ji,}$; $i, j = 1$ and 2.

It was shown [11] that the funnel plot of the averages of bits generated by a Markovian source at equal self-transition probabilities $p_{11} = p_{00} = p > 0{,}5$ will be symmetrically broadened about 50%, thus replicating the funnel plot of IMMI data free of biases. Such self-transition probabilities imply persistence of binary states in the Markovian bit sequence. The mathematical treatment supporting the Markovian model has been presented elsewhere [12-13]. The correctness of the Markovian model is confirmed through computer generated funnel plots of Markovian data obtained at prescribed self-transition probabilities, $p_{ij}$ to which the confidence interval curves are added through equations (6), (7) and (8).

Examples of funnel plots of simulated Markovian data are shown in figure 3. It is noticeable that whenever the self-transition probabilities are equal, even if different than 50%, then the funnel plots converge to $\wp = 50\%$, the most representative effect size in the database, as described in equation (8a). In that sense the Markovian process is able to mimick a random process as far as the statistical average is concerned.

The confidence intervals of Markovian data will now follow the more general equation

$$\text{pi} = \wp \pm z_0 \cdot \sigma_0 \cdot V = \wp \pm z_0 \cdot \sqrt{\frac{\wp(1-\wp)}{N}} \cdot V \qquad (6)$$

considering also equation (8b), from which the size of study N is estimated as a function of pi, for values of $z_0$ as in equations (4) and (5)

$$N = \frac{z_0^2 \cdot V^2 \cdot [\wp \cdot (1-\wp)]}{(\text{pi}-\wp)^2} \qquad (7)$$

where $\wp \overset{p_{11}=p_{00}}{=} 0{,}5$ (figure 3a, b and c). It must be emphasized that p in equation (8c) is the Markovian self-transition probability in this two-state process and must not be confused with the theoretically expected probability of hits, that coincides with the most representative value of pi on the funnel plot in relatively large databases, $\wp$. The replication of Markovian data at unequal self-transition probabilities $p_{11} \neq p_{00}$, $V \neq 1$, $\wp \neq 50\%$, is shown in figure 3d.

$$\begin{aligned} \wp &= \frac{1-p_{00}}{2-(p_{11}+p_{00})} \overset{p_{11}=p_{00}}{=} 0{,}5 && \text{(a)} \\ \sigma_0 &= \sqrt{\frac{\wp(1-\wp)}{N}} \overset{p_{11}=p_{00}}{=} \frac{0{,}5}{\sqrt{N}} && \text{(b)} \\ V &= \sqrt{\frac{p_{11}+p_{00}}{2-(p_{11}+p_{00})}} \overset{p_{11}=p_{00}=p}{=} \sqrt{\frac{p}{1-p}} && \text{(c)} \end{aligned} \qquad (8)$$

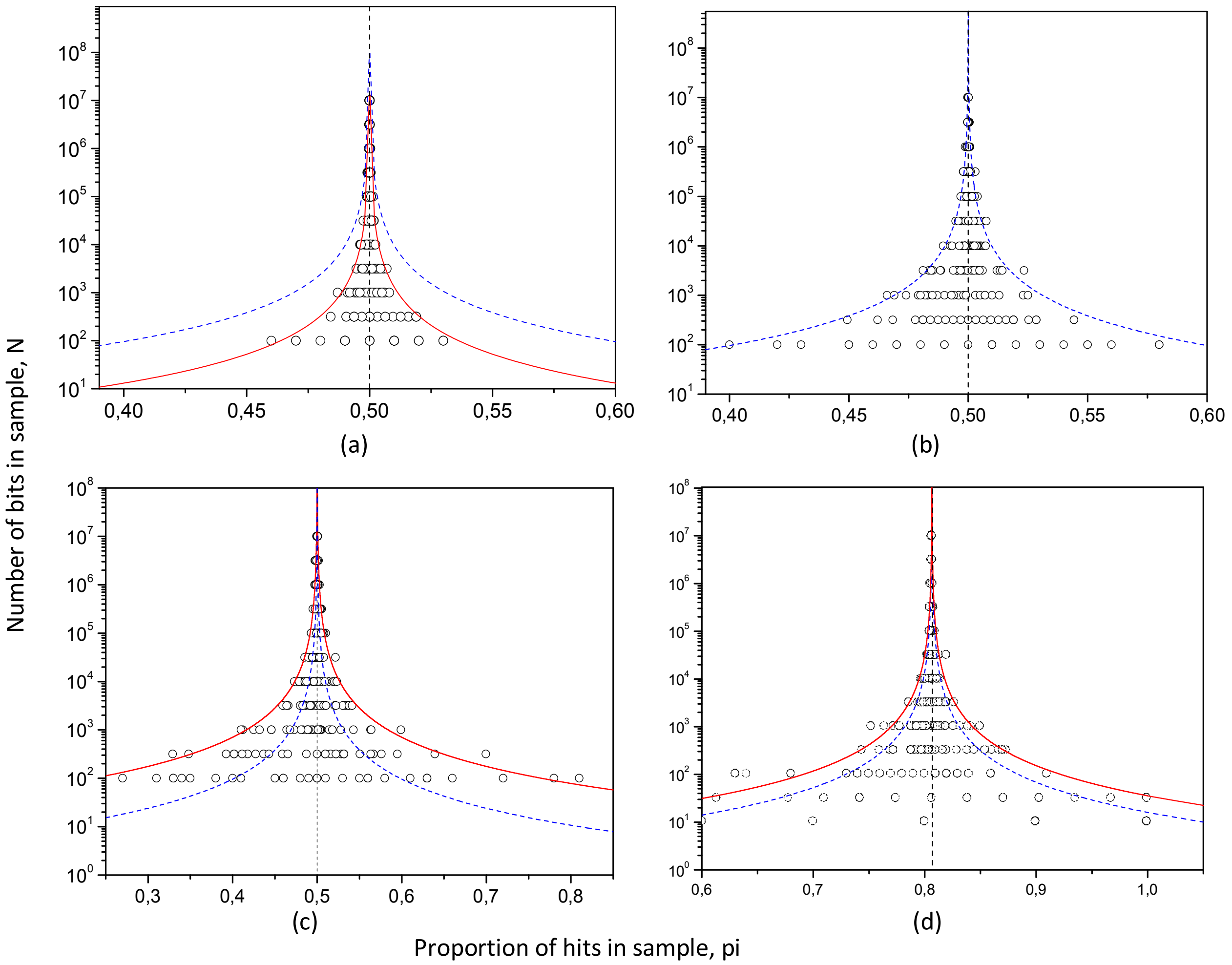

Fig. 3. Examples of funnel plots of simulated Markovian bit sequences generated at self-transition probabilities, $p_{11}$ and $p_{00}$. Open circles: The average of the proportion of 1's in 30 such samples of bits of size N. Vertical dashed line: The position $\wp$ of the most representative proportion of 1's, equation (8a); Red solid lines: 95% confidence interval of biased Markovian data, $V \neq 1$, equations (6), (7) and (8c); Blue dashed lines: 95% confidence interval of random data, V=1, equations (6), (7) and (8c). Self-transition probabilities: (a) $p_{11} = p_{00} = 0{,}12$, $\wp = 0{,}5$; $V = 0{,}37$; (b) $p_{11} = p_{00} = 0{,}5$; $\wp = 0{,}5$, $V = 1$; (c) $p_{11} = p_{00} = 0{,}88$, $\wp = 0{,}5$; $V = 2{,}71$; (d) $p_{11} = 0{,}88$ and $p_{00} = 0{,}5$, $\wp = 0{,}807$; $V = 1{,}49$.

The variance factor V, table 2, is graphically estimated by fitting the confidence interval of equation (7) on the funnel plot for $z_0 = 1{,}96$ (statistical level 95%) and allowing for the variance modulation. By introducing the graphically estimated V in equation (8c), the self-transition pobability is obtained, which for the IMMI database is $p = V^2/(V^2+1) = 0{,}83$ , table 2.

The IMMI records can be arranged to form a time series, assuming similar arrangement of the associated bits in the studies. The correlation coefficient $C_1$ between two such adjacent true IMMI RNG bits, $b_n$ and $b_{n+1}$; n =1,2,…,N-1 in a time series of N such bits has been estimated by [14, 15],

$$C_1 = \langle b_n b_{n+1} \rangle = 2p - 1 \qquad (9)$$

considering that

$p_{ij}$; $i, j = 0,\ 1$ and that $\sum p_{ij} = 1$, since for the product:

$$b_n * b_{n+1} = \begin{cases} 1: & 1 \to 1 \\ 0: & \text{otherwise} \end{cases}$$

The correlation coefficient between two adjacent bits generated by the hypothetical true RNG IMMI Markovian process, for $p = 0{,}83$ is of moderate magnitude $C_1 = 0{,}66$, equation (9) and table 2. When neighbours in the Markovian bit sequence are separated by a distance of k steps, their correlation coefficient decreases in magnitude according to

$$C_k = (2p-1)^k \qquad (10)$$

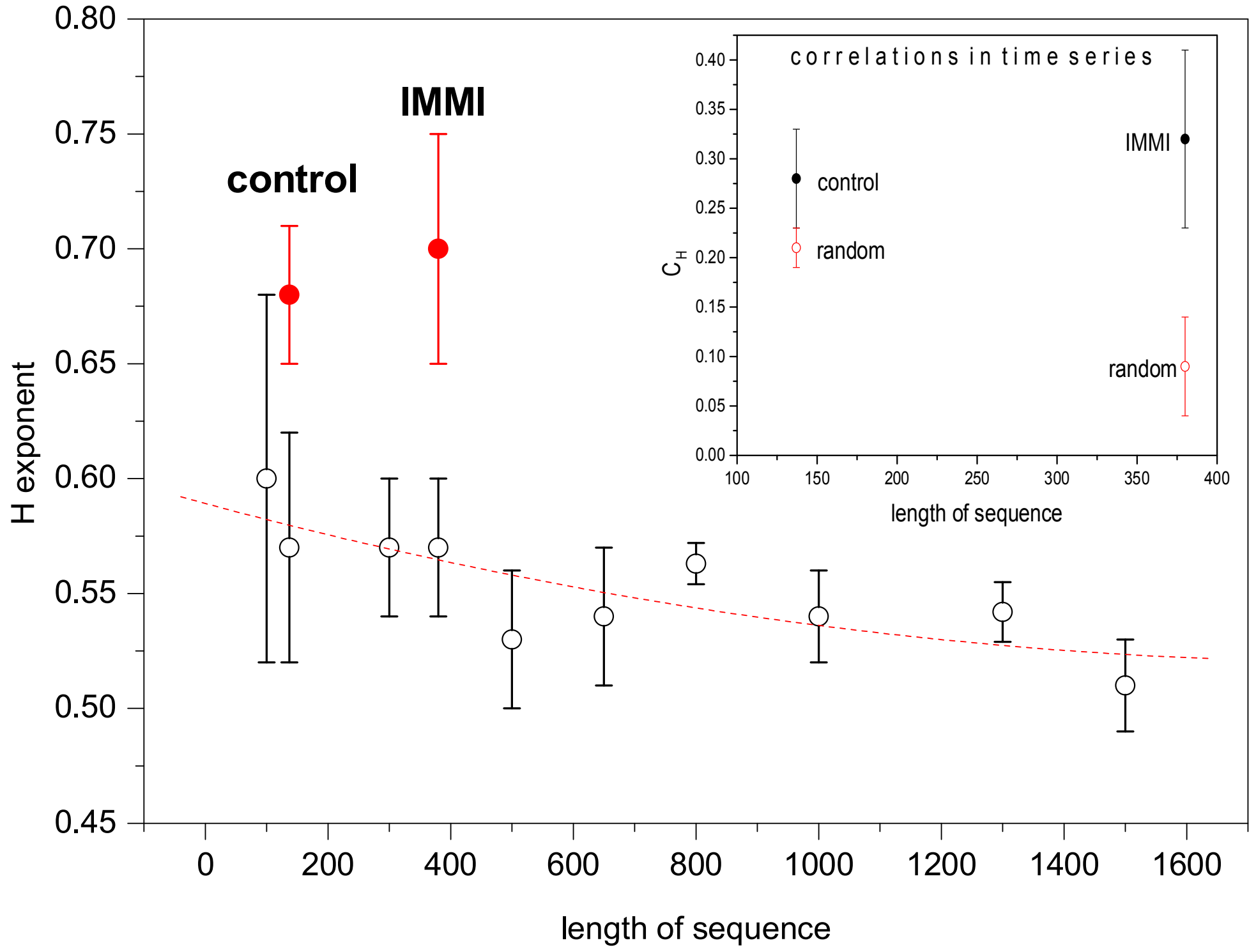

Fig. 4. Hurst exponents of time series of true RNG data, as well as of random data generated online [22], estimated through the rescaled range analysis, equ. (13). Full red circles: IMMI and control true RNG time series, as indicated. Open circles: Hurst exponents of random data sequences generated online. Dotted line: 2nd order polynomial fit to the Hurst exponents of the random data sequences. Inset: Coefficient $C_H$ of long-range correlations, equ. (14), in the IMMI and control RNG time series, as well as in the time series of random data. See table 3 and text for more details.

Any correlations between neighbours separated by a distance as large as k~30 have already significantly weakened to practically zero.

It is shown that the putative IMMI mechanism, seen as a Markovian source, generates correlated bits due to the persistence of the same binary state regardless of state, manifested at 83% self-transition probability. In other words, the persistence of the same IMMI binary outcome is the same both in the direction of intention as well as against it. This result is bizarre, considering the IMMI hypothesis, implying possible psychological factors involved during the related experiments.

The direct consequence of such bi-directional correlations among bits is the broadening of the funnel plot of IMMI data. Unfortunately, information about the arrangement of bits in experiments collected over the last 35 years is lacking for practical reasons. The estimated correlation coefficient between first-neighbour Markovian IMMI bits, (66%) table 2, stands only as an indicator of a correlating process spread on average across experiments. In the next paragraph the nature of this correlating process will be investigated.

### *The nature of the correlations in IMMI RNG database*

*Rescaled range analysis*

To investigate the nature of correlations in the IMMI database, the reported true RNG effect sizes were arranged into time series (from year 1969 to year 2004) and subjected to the rescaled range analysis, (R/S). The arrangement of data into time series was performed in terms of year of publication as well as month of publication within the same year –whenever possible– on the basis of the available information. To apply the R/S analysis, the records $x_i, i = 1,2,3\ldots, N$ in the time series first go through the linear transformation of equation (11)

$$X(t,N)=\sum_i^t\left(x_i-\bar{x}_N\right)=\sum_i^t x_i - t\bar{x}_N,\; i=1,2,3,\ldots,N \quad (11)$$

generating the time series of the rescaled records, $X(t,N)$ [16]. The range of the rescaled variable $X_i$, defined as

$$R(N)=\max X(t,N)-\min X(t,N),\; 1\le t\le N \quad (12)$$

is then divided by the standard deviation, S, of records $x_i$ to yield the rescaled range, R/S. It was shown that the R/S variable follows the exponential law [17]

$$R/S \propto N^H \quad (13)$$

where H is known as the Hurst exponent, shown to be equal to 0,5 for sequences of true random records. The rescaled range analysis, applied on the time series of true IMMI RNG data by use of computer software [17], shows the presence of persistent correlations, $0{,}5 < \mathrm{H} < 1$, table 3 and figure 4.

It was shown that the past and future increments in the time series of $x_i$, considered to be stationary and having zero mean exhibit persistent correlations for values of the Hurst exponent ranging as $0{,}5 < \mathrm{H} < 1$ [18, 19]. It was also shown that the correlation coefficient, $C_H$, between past and future increments of the accumulated departure from the mean in the time series depends on H [20, 21] according to

$$C_H = 2^{2H-1} - 1 \qquad (14)$$

The H exponent and therefore the extent of correlations in the true RNG time series of effect sizes depends on the length N of time series, figure 4. Because the IMMI and control true RNG databases have different size, assessing the extent of correlations in the IMMI time series against that of control data, requires a reference time series of random data comparable in form and length of time series.

For that purpose, ten sequences of random numbers equivalent in form to the reported effect sizes, –i.e. in z-score format estimated by equation (2)– and of appropriate size, were generated online [22] and their average H exponent was used as reference. Compared with the random sequences, the IMMI time series are shown to exhibit distinct persistent correlations, where as the correlations in the control time series are comparable in magnitude and sign to those present in the random sequences, within statistical error.

To appreciate, however, that the correlations observed in the time series of true RNG data are due to the sequential ordering imposed by the data arrangement per date of publication, each of the two true RNG time series was subjected to ten randomizations. The average Hurst exponents from the ten randomised sequences and the corresponding average correlation coefficients were thus obtained, table 3. These parameters become now comparable in magnitude with those obtained from the random sequences [22]. The persistence of like scores arranged per date of publication has been destroyed. Its presence was not a random event. The observed weak persistent correlations in the time series of control data (comparable to those in random sequences), are due to the high level of publication bias in the database and its smaller size.

## RESULTS AND DISCUSSION

### *Statistical quality of the RNG databases*

Two databases of experimental records testing the behaviour of true random number generators (RNG) operating under experimenter-defined conditions are investigated here. The first condition makes reference to direct human intention to ostensibly influence the concurrent outcomes of the RNG process, designated the IMMI data. The other condition is set to contrast the previous one, designated the control data. In the latter case, data are collected in order to test the appropriateness of the RNG devices in their function as true stochastic processes.

Examining the statistical quality of the two true RNG databases through their funnel plots, the strong presence of publication bias becomes evident becomes evident, especially at regions of small size studies. The publication bias interestingly appears in respective funnel plots over opposite regions where either the RNG records refute the tested hypothesis, or where it is falsely considered that they would refute the tested hypothesis. For instance, in the IMMI database the publication bias makes its presence at scores below 50%, whereas the same kind of bias in the control database appears at scores above 50%. Statistical estimates of the average effect size in the database will be, therefore, flawed. The estimated effect size will be either shifted above or below the value observed on the funnel plot to be the most representative effect size in both RNG databases, i.e. $\wp = 50\%$, resulting in false statistical evidence of the tested hypothesis.

As the IMMI and control databases consist of data generated by the same RNG processes, merging the two makes sense. The statistical power of the larger database has increased, while the statistical average of all effect sizes in the merged database is rendered to the value expected by a true unbiased stochastic process; 50%. The opposing contributions of publication bias in the two databases are counterbalanced and thus negated.

In spite of the confusing presence of publication bias in both databases, the anticipated symmetric shape about a 50% effect size is clearly highlighted in both true RNG funnel plots. The IMMI hypothesis posits that the overall statistical average of effect sizes will be shifted to above 50%. In that sense, the evidence provided by the funnel plot of IMMI data has rejected the tested IMMI hypothesis: The statistical outcomes of the RNG devices have not been modulated by the endeavour of direct mental interferences.

## *Correlations in the IMMI RNG database*

Persistent long-range correlations were observed in the time series of IMMI experimental records of moderate magnitude (correlation coefficient = 32%), although these records were supposedly published by independent experiments. Such correlations underline a tendency to keep persistently the reported records away from the chance statistical average, 50%, i.e. either above as well as below it.

In addition, the successful replication of the funnel plot of IMMI data by the Markovian model (operating at fixed and equal self-transition probabilities) could be thought to suggest the potential presence of an IMMI RNG universal mechanism correlating the binary records upon their generation. Such universal mechanism does not exist, however, as discussed in the previous paragraph, as not being supported by the evidence.

Furthermore, such universal IMMI mechanism could not be responsible for the persistent correlations observed in the time series of IMMI experimental records. The reason is that Markovian correlations are known to die away quickly when averages within batches of the data form new time series [23]. The correlations between Markovian bits would have too short a correlation length (not longer than k=30 neighbours) to have surfaced at the level of experimental records.

To further test this simple understanding, sequences of Markovian bits generated at 83% self-transition probabilities were generated through computer software [24]. They were then divided into batches of variable size and the statistical average of bits in the batches made a new sequence of means that was subjected to the R/S analysis. It was thus practically confirmed that the observed correlations in bit sequences would die away fast in the time series created by the means of such batches of bits.

The nature of persistent long-range correlations binding experimental IMMI effect sizes should be sought in the psychology of experimenters. It is not that they originate from correlations in Markovian bit sequences generated by a universal IMMI mechanism, but the other way around; The long-range correlations present in the sequences of IMMI experimental records have their equivalent at the level of Markovian IMMI bits. This equivalent is expressed as an approximate average preference of 83% for generating the same Markovian bit across experiments, which leads to an approximate average correlation coefficient of 66% between IMMI bits.

## *The experimenter expectancy effect*

The only reasonable explanation for the presence of persistent correlations in the IMMI experimental records, to account for the bi-directionality of the equivalent Markovian correlations at the bit level and the evidence from their funnel plots that rejects the tested hypothesis, is the so-called "experimenter expectancy effect" [25]. The effect refers to the unintended influence of the experimenters' hypotheses or expectations on the results of their research. The experimenter expectancy effect is considered as one of the sources of artifact or error in scientific inquiry. It manifests often in smaller size RNG studies, or RNG studies of slower bit generation, where more handling of data is involved.

In that sense, the outcome of a previously published study will influence the experimenter into unconsciously disregarding present experimental errors while collecting or treating IMMI RNG bits, or into not adequately scrutinizing the treatment of collected bits. Such conduct appears to introduce persistent correlations in the true RNG time-series of effect sizes. Naturally, an experimental result in support of the investigated hypothesis will be reported easier if a preceding publication has reported similar results. Even more so, experimental results that refute the investigated hypothesis will be reported with less hesitation, provided another preceding study has come to a similar negative result.

If persistent correlations appear due to human intervention in handling data in time series of IMMI records where bits are generated at low or medium speed, how are such errors introduced when the RNG processes generate bits at very fast rates? In such cases, opportunities to introduce errors due to data handling must be scarcer yet potentially present, as explained further.

There are instances where RNG data sequences generated during many separate tests need to be merged into the final sequence that represents the overall result of the study. There may be more such sequences to be merged for the final IMMI database, than are sequences merged, for instance, in the case of the control database. In that sense, the making of an IMMI database from very fast RNG processes will require more data handling than the making of the control database.

Control bit sequences are often not generated right after each separate IMMI test, but collectively at the end of a day with an equal number of bits as the total of IMMI bits collected during the same day. Consequently, there will be fewer and longer control sequences merged in order to make the final control database. In other words, there will be fewer chances of data handling or human intervention involved in collecting control data than there is for the IMMI data.

Very fast true RNG devices require to be tested for their proper performance. This is achieved by generating the so-called "calibration" data sequence. This is a much longer bit sequence generated over very long uninterrupted periods. To get one such calibration data sequence, therefore, minimum or zero human intervention is involved.

Confirmation of the above description is provided by a rescaled range analysis has been applied on IMMI, control and calibration sequences that were generated as part of the data reported in [2]. The Hurst exponents were estimated through equation (13). Weak persistent correlations were observed in IMMI data sequences ($H_{IMMI} = 0{,}521$, se = 0,004), less prominent persistent correlations were observed in control data sequences ($H_{control} = 0{,}508$, se = 0,003) and no correlations at all were present in a long calibration data sequence ($H_{calibration} = 0{,}505$, se = 0,006) of comparable length (500.000 records) to that of the IMMI and control data sequences (450.000 records each) [21, 26].

If the robust information published in the true RNG meta-analysis [1] had not been available, such an isolated result above from a part of the data published in one study, may have supported an alternative, yet false, interpretation of the IMMI hypothesis, even though the hypothesis itself concerning overall mean shifts had already been refuted [2].

## CONCLUSIONS

The present analysis applies to all available data that have been collected over a period 35 years from studies examining the question whether direct human intention can modulate the outcomes of true random number generators. Reviving the investigation was considered necessary in view of the many ambiguities surrounding the related evidence to date, taking also into account the scientific and technological implications of such prospect.

After thorough examinination of the amassed related evidence, it is concluded that there is no scientific evidence to support the aforesaid claims. Direct human intention cannot modulate a stochastic process, such as the statistical behaviour of a random number generator. Comparable achievements have been observed only through technological advances of neurosciences in the form of brain-machine interface devices.

The present analysis shows that whatever indications may exist in favour of the notoriously elusive IMMI phenomena, these have clearly arisen from unconscious human intervention on the outcomes of the stochastic process; that is, through direct physical rather than through direct mental processes.

**Acknowledgements** The information regarding the data of the IMMI meta-analysis presented here was kindly provided by Dr. F. Steinkamp, co-author of this meta-analysis.